\documentclass[twocolumn,reprint,aps,floatfix,dvipdfmx]{revtex4-1}
\usepackage{amsmath,amssymb}
\usepackage{graphicx}
\usepackage{bm}
\usepackage{physics}
\usepackage{color}
\usepackage{dcolumn}
\usepackage{ulem}

\begin{document}

\title{Light-induced switching of magnetic order in the anisotropic triangular-lattice Hubbard model}

\author{Hayato Kobayashi$^1$}
\author{Ryo Fujiuchi$^1$}
\author{Koudai Sugimoto$^2$}
\author{Yukinori Ohta$^1$}
\affiliation{
$^1$Department of Physics, Chiba University, Chiba 263-8522, Japan\\
$^2$Department of Physics, Keio University, Yokohama 223-8522, Japan
}

\date{\today}

\begin{abstract}
The time-dependent exact-diagonalization method is used to study the light-induced
phase transition of magnetic orders in the anisotropic triangular-lattice Hubbard model.
Calculating the spin correlation function, we confirm that the phase transition from the
120$^{\circ}$ order to the N\'{e}el order can take place due to high-frequency periodic
fields.  We show that the effective Heisenberg-model Hamiltonian derived from the
high-frequency expansion by the Floquet theory describes the present system very well
and that the ratio of the exchange interactions expressed in terms of the frequency and
amplitude of the external field determines the type of the magnetic orders.  Our results
demonstrate the controllability of the magnetic orders by tuning the external field.
\end{abstract}

\maketitle


One of the most significant themes in condensed matter physics concerns the presence of various long-range orders in correlated electron systems.
In particular, by intense laser-pulse irradiation, which leads the systems to nonequilibrium states, the creation and control of long-range orders have recently been made feasible in experiments.
Typical examples of such attempts include possible photoinduced superconductivity in cuprates \cite{Fausti2011Science, Hu2014NM}, alkali-doped fullerides \cite{Mitrano2016Nature}, FeSe \cite{Suzuki2019CP}, and organic salts \cite{Buzzi2020PRX}, light-induced charge-density waves in LaTe$_3$ \cite{Kogar2020NP}, and band gap control in excitonic insulators Ta$_2$NiSe$_5$ via photoexcitation \cite{Mor2016PRL,
Okazaki2018NC}.

In the manipulation of nonequilibrium states, the concept of ``Floquet engineering" attracts particular attention \cite{Sato2016PRL, Oka2019ARCMP}, where we create a nonequilibrium steady state by applying a time-periodic external field and change the state to the desired one by tuning the amplitude and frequency of the field.
The long-range orders in correlated electron systems can be controlled in this manner.
In the Hubbard-model Hamiltonian, in particular, the electric field by light irradiation is introduced via the Peierls phase substitution into the hopping integrals.
The exchange interactions of the system under light irradiation can thus be derived in the strong-coupling limit of the Hamiltonian  \cite{Mentink2015NC, Kitamura2016PRB}, which controls the phase of the system accordingly.
Theoretical studies made so far include the phase transition from antiferromagnetic to ferromagnetic order \cite{Mentink2015NC, Dasari2019PRB}, switching of superconductivity and the charge-density wave in an attractive Hubbard model \cite{Kitamura2016PRB, Sentef2017PRL, Fujiuchi2020PRB}, etc.
Related experiments have also been carried out in ultracold atom systems \cite{Eckardt2017RMP}.
The success of controlling the exchange interactions has been reported as well in iron oxides \cite{Mikhaylovskiy2015NC}.

In this Letter, motivated by such developments in the field, we focus on the Hubbard model at half-filling defined on the anisotropic triangular lattice (ATL), which is one of the representative systems with geometrical frustration.
Since the frustrated systems have many competing orders in their ground states, we can expect to realize the control and switching of the orders by an external field \cite{Stepanov2017PRL, Kitamura2017PRB, Claassen2017NC, Takasan2019PRB,Jana2020PRB, Bittner2020PRB}.
The ground-state phase diagram of our model has so far been investigated well by numerical approaches, where we know that the N\'{e}el order, $120^{\circ}$ order, and collinear order compete with each other
\cite{Mizusaki2006PRB, Yu2010PRB, Yamada2014PRB, Laubach2015PRB, Misumi2016JPSJ}.
Materials described by this model include inorganic compounds Cs$_2$CuCl$_4$ \cite{Coldea2002PRL} and Cs$_2$CuBr$_4$ \cite{Ono2003PRB} as well as organic compounds such as $\kappa$-(BEDT-TTF)$_2$Cu[N(CN)$_2$]Cl \cite{Lefebvre2000PRL}
and $\kappa$-(BEDT-TTF)$_2$Cu$_2$(CN)$_2$] \cite{Shimizu2003PRL}.

In what follows, we will first prepare the ATL Hubbard Hamiltonian whose ground state is of $120^{\circ}$-type antiferromagnetic order, and simulate the change in this quantum state under a time-periodic external field, where we use the time-dependent Lanczos method.
Then, we will show that by tuning the amplitude and frequency of the field, the initial state with the $120^{\circ}$ order can actually be switched to the N\'{e}el order.
This result will be interpreted using the Floquet effective Hamiltonian obtained from the high-frequency expansion of our model.
We will also perform calculations of the quench dynamics of the Heisenberg model obtained in the strong-coupling limit of the Hubbard model, and confirm the validity of our Floquet analysis.
We will thus demonstrate the controllability of the magnetic orders in the frustrated spin system by tuning the external field.



\begin{figure}[t]
  \includegraphics[width=\columnwidth]{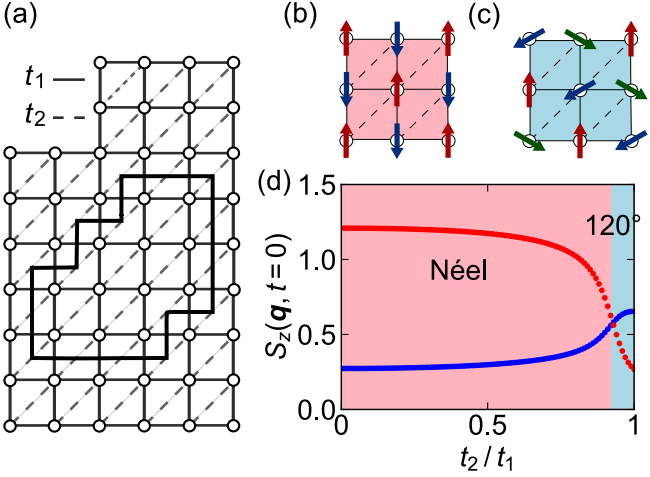}
  \caption{
    (a) Schematic representation of the ATL Hubbard model, where the solid and dashed lines represent the hopping integrals $t_1$ and $t_2$, respectively.
    The area surrounded by the thick solid line indicates the 12-site cluster used in our Lanczos calculations.
    (b), (c) Schematic illustrations of (b) the N\'{e}el order and (c) $120^\circ$ order.
    (d) Spin correlation function $S_z({\bm q},t=0)$ calculated for the ground state with the N\'{e}el order [${\bm q}=(\pi,\pi)$, red line] and $120^\circ$ order [${\bm q}=(2\pi/3,2\pi/3)$, blue line], where we assume $U/t_1=20$.
  }
  \label{fig1}
\end{figure}

The ATL Hubbard model is defined by the Hamiltonian [see Fig.~\ref{fig1}(a)]
\begin{align}
  \hat{\mathcal{H}}
    = -t_1\sum_{\langle i,j \rangle, s} \hat{c}_{i, s}^{\dagger}\hat{c}_{j, s}
      - t_2 \sum_{\langle\langle i,j \rangle\rangle, s} \hat{c}_{i, s}^{\dagger}\hat{c}_{j, s}
      + U\sum_{i}\hat{n}_{i, \uparrow}\hat{n}_{i, \downarrow},
\label{eq:Hubbard-model}
\end{align}
where $\hat{c}^{\qty(\dagger)}_{i, s}$ is the annihilation (creation) operator of an electron at site $i$ with spin $s$, and $\hat{n}_{i, s} = \hat{c}^{\dagger}_{i, s} \hat{c}_{i, s}$ is the electron density operator.
$t_1$ and $t_2$ are the nearest-neighbor (NN) and next-nearest-neighbor (NNN) hopping integrals, respectively, and $U$ is the on-site Coulomb interaction.
The notations $\langle i,j \rangle$ and $\langle\langle i,j \rangle\rangle$ represent the pairs of the NN and NNN sites, respectively.
In the strong-coupling limit $U/t_1\rightarrow\infty$ at half filling, the Hubbard model in Eq.~(\ref{eq:Hubbard-model}) is mapped onto the antiferromagnetic Heisenberg model
defined by
\begin{align}
  \hat{\mathcal{H}}_{\text{eff}}
    = J_1 \sum_{\langle i,j \rangle} \hat{\bm S}_{i}\cdot \hat{\bm S}_{j}
    + J_2 \sum_{\langle\langle i,j \rangle\rangle} \hat{\bm S}_{i}\cdot \hat{\bm S}_{j},
\end{align}
where $\hat{\bm S}_i$ is the spin-1/2 operator at site $i$, and $J_1=4t_1^2/U$ and $J_2=4t_2^2/U$ are the NN and NNN exchange interactions, respectively.
The ground state of this model is of N\'{e}el type [Fig.~\ref{fig1}(b)] at $J_2 / J_1 < 0.83$, which switches to $120^\circ$ type [Fig.~\ref{fig1}(c)] at $J_2 / J_1 > 0.83$ \cite{Weihong1999PRB, Yunoki2006PRB}.

The time-dependent external field is introduced via the Peierls phase.
Then, the hopping integrals are modified as
\begin{align}
  t_{n} \hat{c}_{i, \sigma}^{\dagger} \hat{c}_{j, s}
    \rightarrow
  t_{n} e^{-i\bm A(t) \vdot (\bm r_i - \bm r_j)} \hat{c}_{i, \sigma}^{\dagger} \hat{c}_{j, s},
\end{align}
where $\bm{A} (t)$ is the vector potential at time $t$.
In the present study, we use the vector potential parallel to the NNN direction, i.e.,
$\bm{A} (t) = \frac{1}{\sqrt{2}}\qty[A(t), A(t)]$, where
\begin{align}
  A(t)=
  \begin{cases}
    A_0 e^{-(t-t_0)^2/2\sigma_{\mathrm{p}}^2} \cos \qty[\omega_{\mathrm{p}} \qty(t-t_0)] & (t \leq t_0)
    \\
    A_0 \cos \qty[\omega_p \qty(t-t_0)] & (t > t_0)
  \end{cases}
\end{align}
with the amplitude $A_0$ and frequency $\omega_{\mathrm{p}}$.
In the following, we assume $\sigma_\mathrm{p}=2.0/t_1$ and $t_0=10/t_1$.
We set the Planck constant $\hbar$, speed of light $c$, elementary charge $e$, and lattice constant to be unity.


Since the Hamiltonian depends on time in the presence of an external field, we need to solve the Schr\"{o}dinger equation to obtain the time evolution of the wave function.
We employ the time-dependent Lanczos method \cite{Park1986JCP, Mohankumar2006CPC} for this purpose.
The time evolution with a time step $\delta t$ is calculated in the corresponding Krylov subspace generated by $M_{\mathrm{L}}$ Lanczos iterations.
We use the 12-site cluster illustrated in Fig.~\ref{fig1}(a) with periodic boundary conditions, and adopt $\delta t = 0.01 / t_1$ and $M_{\mathrm{L}} = 15$.
As the initial state, we assume $\ket{\psi (t=0)} = \ket{\psi_0}$, where $\ket{\psi_0}$
is the ground state of the Hamiltonian without an external field.

To determine the type of magnetic orders, we calculate the spin correlation function
with momentum $\bm{q}$ at time $t$ written as
\begin{align}
  S_z(\bm q,t)
    = \frac{1}{L} \sum_{i,j}
    S^z_{ij} (t)
      e^{i \bm{q} \vdot \qty(\bm{r}_i-\bm{r}_j)},
\end{align}
where $S^z_{ij} (t) = \mel{\psi(t)}{\hat{S}^z_i \hat{S}^z_j}{\psi(t)}$ is the spin correlation in the real space.
The ordering vector $\bm{q}=\bm{Q}_{\text{N\'{e}el}} = (\pi,\pi)$ and $\bm{q}=\bm{Q}_{120^{\circ}} = (2\pi/3,2\pi/3)$ correspond to the N\'{e}el order and $120^{\circ}$ order, respectively.
The spin correlation functions calculated as a function of $t_2 / t_1$ at $U / t_1 = 20$
are illustrated in Fig.~\ref{fig1}(d), where we clearly find the phase transition from the N\'{e}el order to the $120^\circ$ order at $t_2 / t_1 \simeq 0.93$.
Throughout the main text of this Letter, we use the values $U / t_1 = 20$ and $t_2 / t_1 = 0.95$, for which the ground state is the $120^\circ$ order.
Other cases, where the ground state has the the N\'{e}el order and the nearest-neighbor hopping integral satisfies $t_2/t_1=1$, are also discussed in Supplemental Material \cite{SM}.
We also define the time average of the spin correlation function from time $t = t_{\mathrm{i}}$ to $t_{\mathrm{f}}$ as
\begin{equation}
  \bar{S}_z(\bm q) = \frac{1}{t_{\mathrm{f}} - t_{\mathrm{i}}}
  \int^{t_{\mathrm{f}}}_{t_{\mathrm{i}}} \dd{t} S_z (\bm{q}, t)
\end{equation}
and the difference between the time average and initial value as $\Delta S_z (\bm{q}) = \bar{S}_z(\bm q) - S_z (\bm{q}, t = 0)$.


\begin{figure}[t]
  \centering
  \includegraphics[width=\columnwidth]{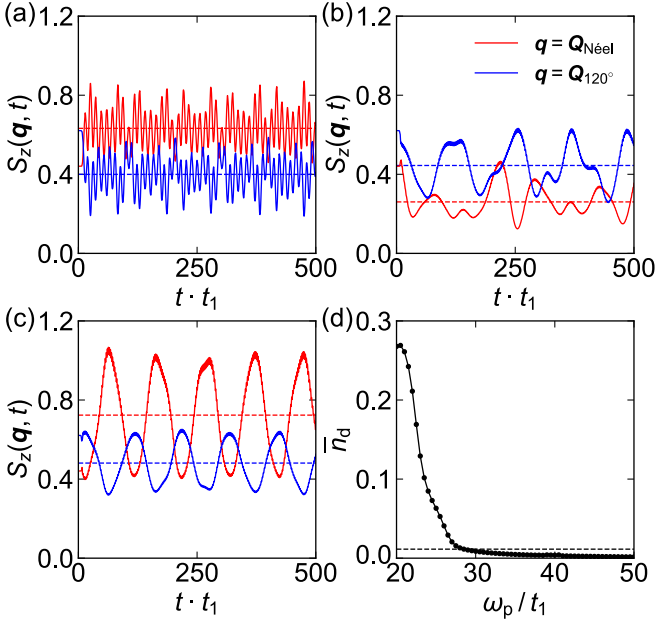}
  \caption{
    Calculated time evolution of the spin correlation function $S_z(\bm{q},t)$ at $\bm{q}=(\pi,\pi)$ (red line) and at $\bm{q}=(2\pi/3,2\pi/3)$ (blue line), where we assume $\omega_{\mathrm{p}}/t_1=30$ with (a) $A_0 = 1.0$, (b) $A_0 = 2.0$, and (c) $A_0 = 4.0$.
    Here, the dashed lines indicate $\bar{S}_z (\bm{q})$ averaged from $t_{\mathrm{i}}=10/t_1$ to $t_{\mathrm{f}}=500/t_1$.
    In (d), we show the calculated double occupancy $\bar{n}_{\mathrm{d}}$ averaged from $t_{\mathrm{i}}=10/t_1$ to $t_{\mathrm{f}}=500/t_1$, assuming $A_0 = 4.0$.
    The dashed line indicates the value of $n_{\mathrm{d}} (t=0)$.
}
\label{fig2}
\end{figure}

The calculated results for the spin correlation function $S_z(\bm{q}, t)$ as a function of time are shown in Figs.~\ref{fig2}(a)-\ref{fig2}(c).
We find that, with the irradiation of light with $A_0 = 1.0$ and $\omega_{\mathrm{p}} = 30 / t_1$, the $120^\circ$ order stable in the ground state is switched to the N\'{e}el order [see Fig.~\ref{fig2}(a)], which clearly indicates a magnetic phase transition induced by an external field.
However, when we increase the amplitude of the external field to $A_0 = 2.0$, the state remains to be of $120^\circ$ order after the light irradiation [see Fig.~\ref{fig2}(a)].
When we continue to increase the amplitude to $A_0 = 4.0$, the system again shows a transition to the N\'{e}el order [see Fig.~\ref{fig2}(c)].
The present results suggest that the light-induced phase transition may occur in the ATL Hubbard model by adjusting the intensity of light.
We note that the possibility of transitions to other types of orders is excluded from the analysis of the real-space spin correlation functions, as discussed in the Supplemental Material \cite{SM}.

\begin{figure}[t]
  \centering
  \includegraphics[scale=0.5]{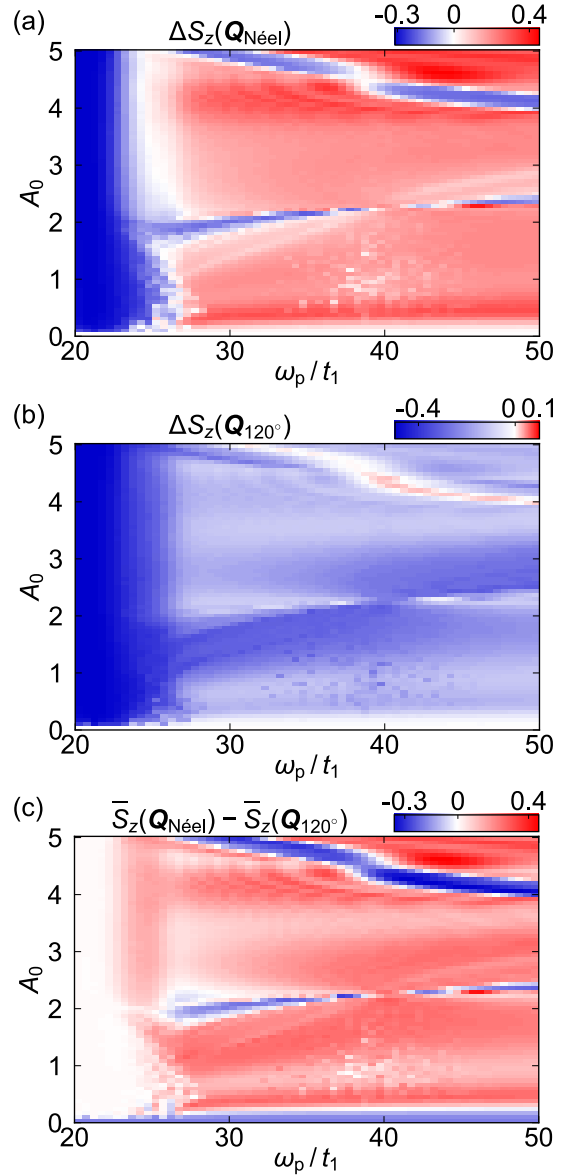}
  \caption{
    Differences in the spin correlation function calculated in the parameter space $(\omega_{\mathrm{p}}/t_1,A_0)$.
    Shown are
    (a) $\Delta S_z (\bm{Q}_{\text{N\'{e}el}})$,
    (b) $\Delta S_z (\bm{Q}_{120^{\circ}})$, and
    (c) $\bar{S}_z (\bm{Q}_{\text{N\'{e}el}}) - \bar{S}_z (\bm{Q}_{120^{\circ}})$.
  }
  \label{fig3}
\end{figure}

To explore the parameter regions where the $120^{\circ}$ order remains or the N\'{e}el order overcomes after the light irradiation, we calculate the difference in the spin correlation function $\Delta S_z (\bm{q})$ in the parameter space $(\omega_{\mathrm{p}}/t_1,A_0)$.
The results are shown in Fig.~\ref{fig3}(a) for the N\'{e}el order and in Fig.~\ref{fig3}(b) for the $120^{\circ}$ order.
These results clearly indicate that when the $120^{\circ}$ order is suppressed, the N\'{e}el order is complementarily enhanced.
In addition, both orders are strongly suppressed at $\omega_{\mathrm{p}} / t_1 < 25$, where we note that the double occupancy defined as
\begin{align}
  n_{\mathrm{d}} (t)
    = \frac{1}{L} \sum_{i} \bra{\psi(t)} \hat{n}_{i, \uparrow} \hat{n}_{i, \downarrow} \ket{\psi(t)}
\end{align}
increases [see Fig.~\ref{fig2}(d)] and therefore the charge excitations occurring across the Mott-Hubbard gap ($\simeq U / t_1$) increase, leading to the suppression of spin fluctuations.
In Fig.~\ref{fig3}(c), we show the calculated result for $\bar{S}_z (\bm{Q}_{\text{N\'{e}el}}) - \bar{S}_z (\bm{Q}_{120^{\circ}})$.
This result indicates which order is realized in the parameter space after the light irradiation; in the red region, the N\'{e}el order appears, while in the blue region, the $120^{\circ}$ order remains.


\begin{figure}[t]
  \centering
  \includegraphics[scale=0.5]{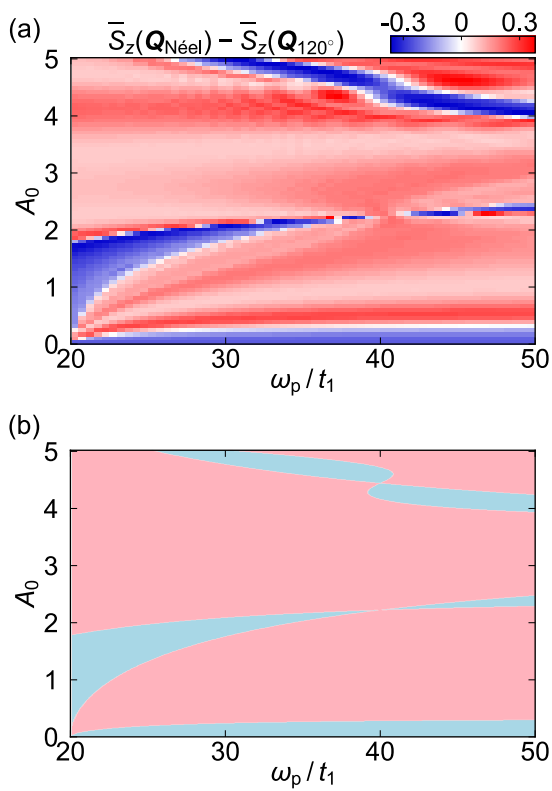}
  \caption{
    (a) Difference $\bar{S}_z (\bm{Q}_{\text{N\'{e}el}}) - \bar{S}_z (\bm{Q}_{120^{\circ}})$ calculated for the effective Heisenberg model Eq.~(\ref{eq:Floquet-effective-Hamiltonian}) in the parameter space $(\omega_{\mathrm{p}}/t_1,A_0)$.
    (b) Ratio of the Floquet effective exchange interactions, where the region of $J_2^{\text{eff}}/J_1^{\text{eff}} < 0.83$ ($J_2^{\text{eff}}/J_1^{\text{eff}} > 0.83$) is indicated by the red (blue) color.
  }
  \label{fig4}
\end{figure}

To discuss the origin of the light-induced phase transition, we analyze the model using the Floquet theory \cite{Mentink2015NC, Kitamura2016PRB}.
Applying the high-frequency expansion to our Hubbard-model Hamiltonian with an external field $A(t) = A_0 \cos \omega_{\mathrm{p}} t$, we obtain the effective Hamiltonian in the strong-coupling limit as \cite{Kitamura2016PRB}
\begin{align}
  \hat{\mathcal{H}_{\text{eff}}}
    = J_1^{\text{eff}} \sum_{\langle i,j \rangle} \hat{\bm S}_{i}\cdot \hat{\bm S}_{j}
      + J_2^{\text{eff}}\sum_{\langle\langle i,j \rangle\rangle} \hat{\bm S}_{i}\cdot \hat{\bm S}_{j},
\label{eq:Floquet-effective-Hamiltonian}
\end{align}
where
\begin{align}
  J_1^{\text{eff}}
    &= \sum_{m = -\infty}^{\infty}(-1)^m \frac{4 t_1^2 \mathcal{J}_m (A_0/\sqrt{2})}{U + m \omega_{\mathrm{p}}}
\intertext{and}
  J_2^{\text{eff}}
    &= \sum_{m = -\infty}^{\infty}(-1)^m \frac{4 t_2^2 \mathcal{J}_m(\sqrt{2}A_0)}{U + m \omega_{\mathrm{p}}}
\end{align}
are the NN and NNN Floquet effective exchange interactions, respectively.
Here, $\mathcal{J}_m (x)$ is the $m$-th Bessel function.
Using this model, we perform the calculation of the quench dynamics \cite{Kawamura2017CPC}, where $J_1$ and $J_2$ are suddenly changed to $J_1^{\text{eff}}$ and $J_2^{\text{eff}}$, respectively, at $t = 0$.
The result for the difference $\bar{S}_z (\bm{Q}_{\text{N\'{e}el}}) - \bar{S}_z (\bm{Q}_{120^{\circ}})$ thus calculated is shown in Fig.~\ref{fig4}(a) in the parameter space $(\omega_{\mathrm{p}}/t_1,A_0)$, which we find is consistent with the result obtained for the Hubbard model at least in the region $\omega_{\mathrm{p}}/t_1\gtrsim 25$ [see Fig.~\ref{fig3}(c)].
The inconsistency found in the region $\omega_{\mathrm{p}}/t_1\lesssim 25$ comes from the enhancement of the double occupancy, which cannot be explained by the strong-coupling expansion.
We thus conclude that in a regime of sufficiently high frequency, the result obtained from the Floquet effective Hamiltonian Eq.~(\ref{eq:Floquet-effective-Hamiltonian}) well explains the behaviors of the Hubbard model in the strong-coupling region under a time-periodic external field.

We also calculate the phase diagram of the effective Hamiltonian simply from the ratio of the effective exchange interactions $J_2^{\text{eff}}/J_1^{\text{eff}}$.
The result is shown in Fig.~\ref{fig4}(b), where the phase boundary is determined as the line $J_2/J_1 = 0.83$, at which the phase transition between the N\'{e}el and $120^{\circ}$ orders occurs in the ground state of the ATL Heisenberg model.
We thus find that the N\'{e}el order is preferred in the red region ($J_2^{\text{eff}} / J_1^{\text{eff}} < 0.83$), while the $120^{\circ}$ order is preferred in the blue region ($J_2^{\text{eff}} / J_1^{\text{eff}} > 0.83$).
We thus clearly find that the phase diagram obtained by the quench-dynamics calculation is consistent with the phase diagram determined from the ratio of the exchange interactions, implying that the magnetic order realized by the light irradiation can be predicted from the Floquet theory.

We note that there are regions where $J^{\mathrm{eff}}_1 < 0$ and $J^{\mathrm{eff}}_2 < 0$, i.e., the regions where the ground state of the Hamiltonian Eq.~(\ref{eq:Floquet-effective-Hamiltonian}) is ferromagnetic.
Our calculated results, however, do not indicate the presence of such regions.
This is because the electric field never flips the spins, or the total spin is conserved by light irradiation.
In addition, since the time evolution of the sign-reversed Hamiltonian is exactly identical with the reversed time evolution of the original Hamiltonian \cite{Mentink2015NC}, we can discuss the dynamics of an effective Hamiltonian with $J^{\mathrm{eff}}_1 < 0$ and $J^{\mathrm{eff}}_2 < 0$ by using the effective Hamiltonian with $J^{\mathrm{eff}}_1 > 0$ and $J^{\mathrm{eff}}_2 > 0$.
We also note that there is a region where $J_1^{\text{eff}} \cdot J_2^{\text{eff}} < 0$.
It is known that the N\'{e}el order is preferred in the case of $J_1^{\text{eff}} > 0$ and $J_2^{\text{eff}} < 0$ \cite{Schmidt2014PRB}, and thus the whole phase diagram can again
be interpreted from the Floquet theory.


In summary, we have investigated the time dependence of the spin correlations of an anisotropic triangular Hubbard model at half filling under a time-periodic external electric field using the time-dependent Lanczos method.
We have shown that the $120^\circ$ order can be switched to the N\'{e}el order by tuning the frequency and amplitude of the external field.
To understand the magnetic phase transition under a periodic field, we have introduced the effective Heisenberg-model Hamiltonian by high-frequency expansion.
The phase diagram obtained from the quench dynamics of this effective model is consistent with the results of our Hubbard-model calculations, which implies that the phase diagram obtained by light irradiation can be interpreted by the Floquet theory.
Thus, the switching of magnetic orders can be realized in a frustrated spin system by tuning the amplitude and frequency of the external field.
We hope that our results will shed some light on the possible realization of the photo-control of magnetic orders in frustrated spin systems.


This work was supported in part by Grants-in-Aid for Scientific Research from JSPS
(Projects No.~JP17K05530, No. JP19J20768, No. JP19K14644, and No. JP20H01849).
R.F.~acknowledges support from the JSPS Research Fellowship for Young Scientists.
We acknowledge the use of open-source software $\mathcal{H}\Phi$ \cite{Kawamura2017CPC}.


\end{document}


\title{Supplemental Material for ``Light-induced switching of magnetic order in the anisotropic triangular-lattice Hubbard model''}

\author{Hayato Kobayashi$^1$}
\author{Ryo Fujiuchi$^1$}
\author{Koudai Sugimoto$^2$}
\author{Yukinori Ohta$^1$}
\affiliation{
$^1$Department of Physics, Chiba University, Chiba 263-8522, Japan\\
$^2$Department of Physics, Keio University, Yokohama 223-8522, Japan
}



\maketitle

\renewcommand{\theequation}{S.\arabic{equation}}
\renewcommand{\thefigure}{S\arabic{figure}}


\section{Magnetic structure in real space}

We have determined the photo-induced magnetic orders based on the Fourier coefficients of the spin correlation functions in the main text.
This section confirms the type of induced magnetic orders by showing
the real-space spin correlation function $S^z_{ij} (t)$.

\begin{figure}[b]
  \centering
  \includegraphics[width=0.8\columnwidth]{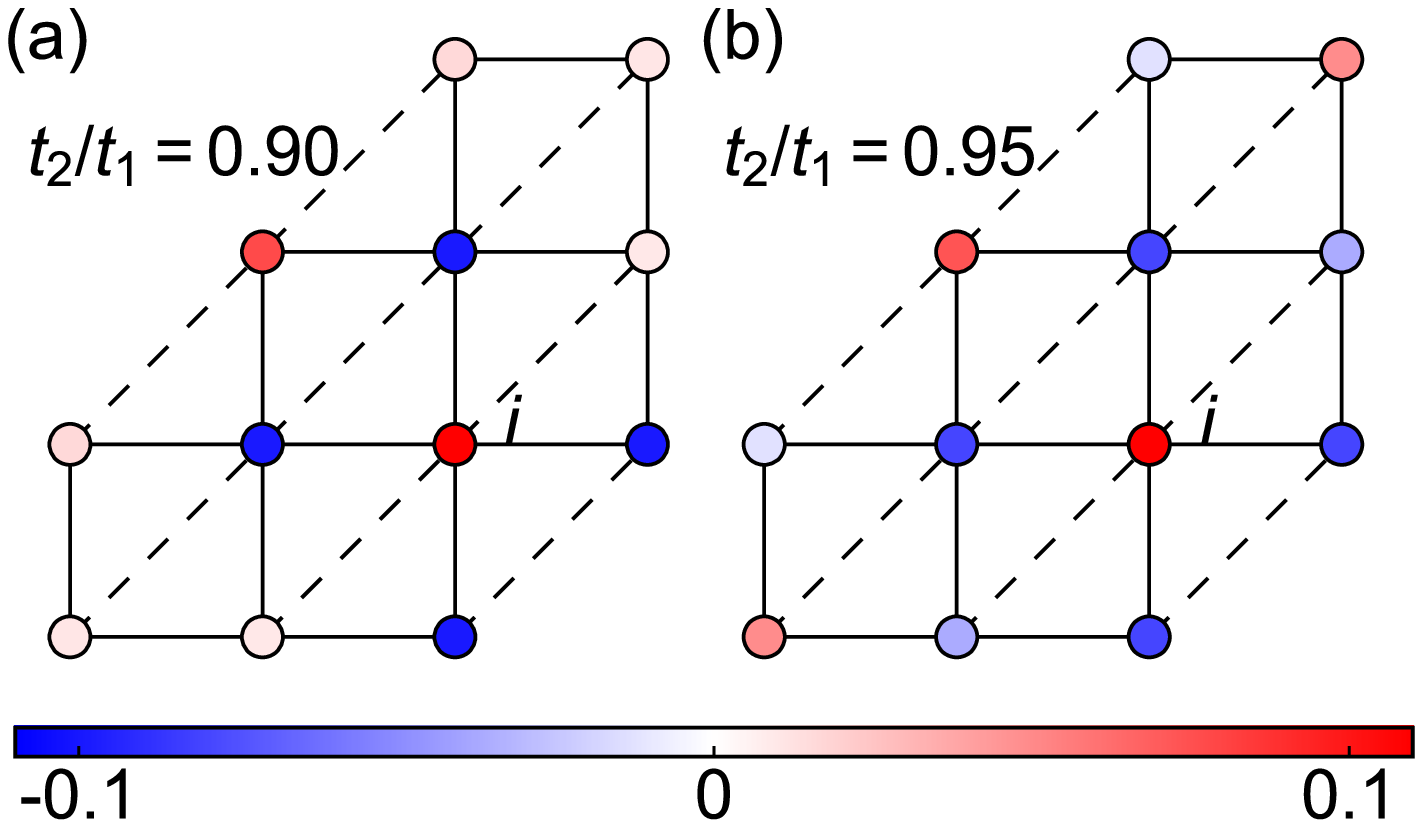}
  \caption{
  Calculated amplitude of the real-space spin correlation functions in the ground state, $S^z_{i,j} (t=0)$.
  The next-nearest-neighbor hopping integral is set to (a) $t_2/t_1 = 0.90$ (N\'{e}el order) and
  (b) $t_2/t_1 = 0.95$ ($120^\circ$ order).
  The value at the origin ($i=j$) is $S_{ii} (0) \simeq 0.244$ in both cases.
  }
  \label{figS1}
\end{figure}

Figures \ref{figS1}(a) and (b) show the real-space spin correlation functions
in the ground state at $t_2 / t_1 = 0.90$ and at $t_2 / t_1 = 0.95$, respectively.
In the case of $t_2 / t_1 = 0.90$, the spin correlation functions at the neighboring sites show the opposite sign, indicating that the ground state is the checkerboard-type
N\'{e}el-order state whose ordering vector is equal to $\bm{q} = (\pi, \pi)$.
On the other hand, in the case of $t_2 / t_1 = 0.95$, the ordering vector of the
spin correlation function changes to $\bm{q} = (2\pi/3, 2\pi/3)$, whose pattern
just corresponds to the $120^\circ$-order state.

\begin{figure}[t]
  \centering
  \includegraphics[width=\columnwidth]{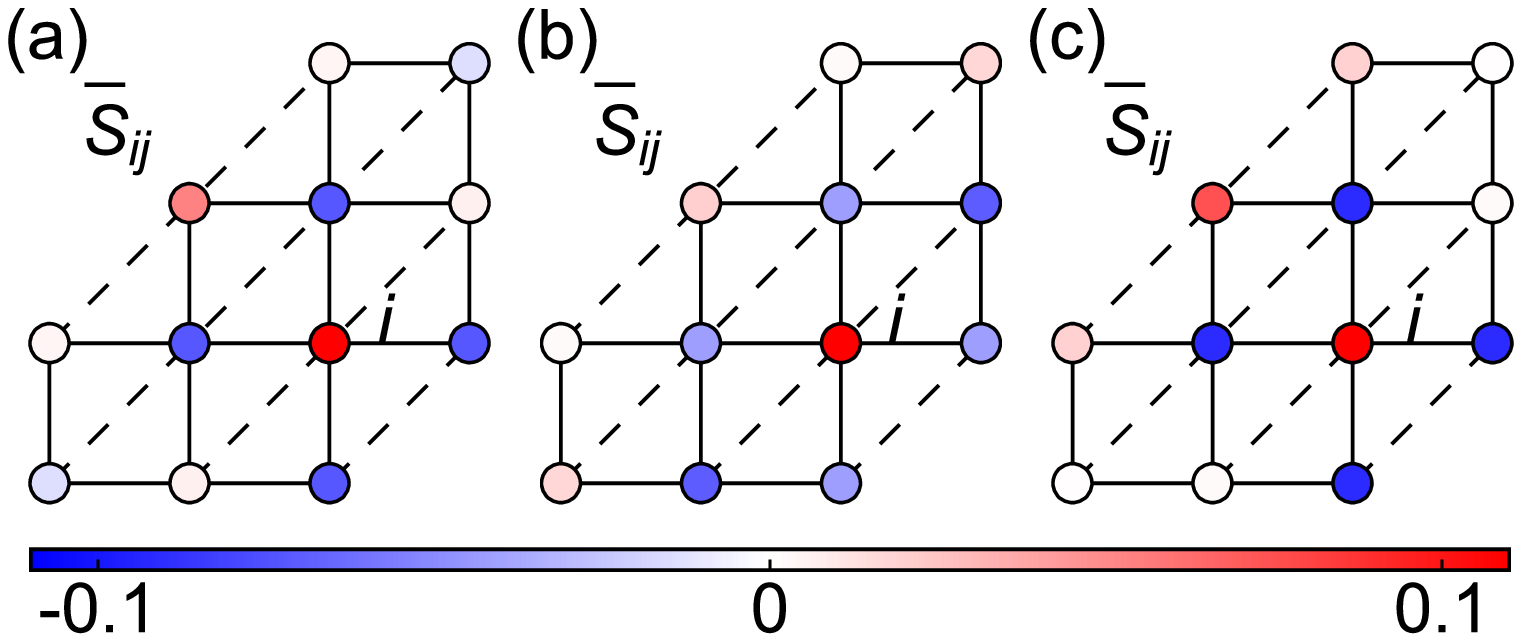}
  \caption{
    Calculated time-averaged amplitude of the real-space spin correlation functions $S^z_{i,j} (t)$ at $t_2 / t_1 = 0.95$.
    We assume $\omega_{\mathrm{p}} = 30$ with (a) $A_0 = 1.0$, (b) $A_0 = 2.0$, and (c) $A_0 = 4.0$.
    The time-averaged values at the origin ($i=j$) are (a) $\bar{S}^z_{ii} \simeq 0.242$,
    (b) $\bar{S}^z_{ii} \simeq 0.243$, and (c) $\bar{S}^z_{ii} \simeq 0.246$.
  }
  \label{figS2}
\end{figure}

Next, we discuss the magnetic order in the photo-excited state at $t_2 / t_1 = 0.95$,
where the ground state is the $120^\circ$ order.
Figure \ref{figS2}(a) is the time-averaged real-space spin correlation function
$\bar{S}^z_{ij}$ at $A_{0} = 1.0$ and $\omega_{\mathrm{p}} = 30$.
We find that the magnetic configuration switches from the $120^\circ$ order to the N\'{e}el order, implying that the magnetic phase transition occurs by light irradiation.
However, as shown in Fig.~\ref{figS2}(b), when the amplitude of the external field is
increased to $A_0 = 2.0$, the magnetic configuration remains to be $120^\circ$ order
even after the photo-irradiation.
With further increasing the amplitude to $A_{0} = 4.0$, the spin configuration again becomes the N\'{e}el order, as shown in
Fig.~\ref{figS2}(c).

We thus conclude that the magnetic structure expected from the real-space spin
correlation functions of the photo-excited states is consistent with the assignments
based on the Fourier component of spin-correlation functions shown in the main text.
Possibilities of other kinds of magnetic orders are therefore excluded.

\section{Transition from N\'{e}el order to 120$^{\circ}$ order}
\label{Sec:Neel-to-120-order}

We discuss the photo-induced magnetic phase transition from the $120^\circ$ order to N\'{e}el order in the main text.
In this section, we consider the opposite case, i.e., the phase transition from the N\'{e}el order to the $120^\circ$ order.

\begin{figure}[tb]
  \centering
  \includegraphics[width=\columnwidth]{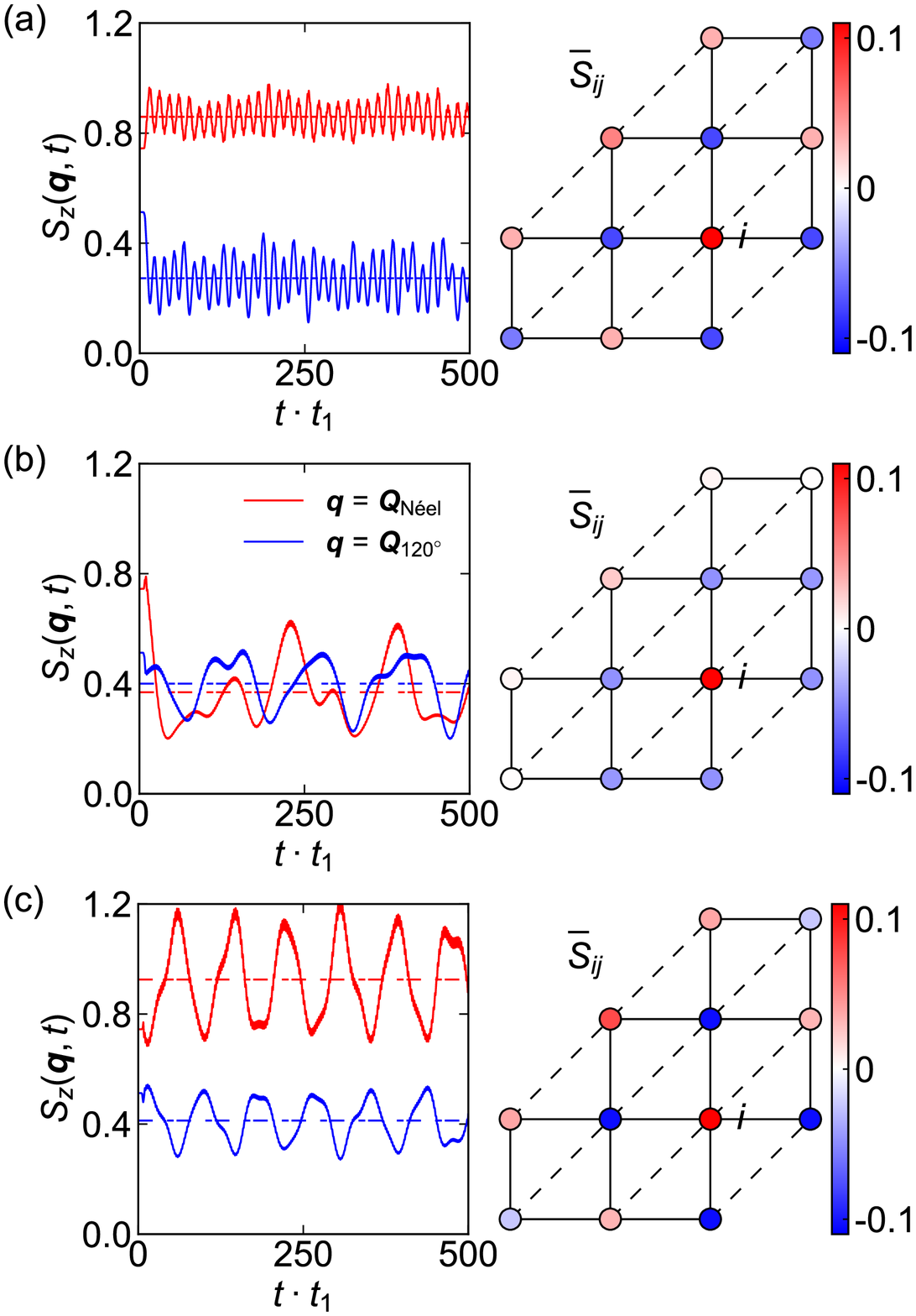}
  \caption{
  Left panels: calculated time evolution of the spin correlation function $S_z(\bm{q},t)$ at
  $\bm{q}=(\pi,\pi)$ (red line) and at $\bm{q}=(2\pi/3,2\pi/3)$ (blue line), where the dashed
  lines indicate $\bar{S}_z (\bm{q})$ averaged from $t_{\mathrm{i}}=10/t_1$ to $t_{\mathrm{f}}=500/t_1$.
  Right panels: calculated time-averaged real-space spin correlation functions $\bar{S}^z_{ij}$.
  We assume  $\omega_{\mathrm{p}}/t_1=30$ with
  (a) $A_0 = 1.0$,
  (b) $A_0 = 2.0$, and
  (c) $A_0 = 4.0$.
  The time-averaged values at the origin ($i=j$) are (a) $\bar{S}^z_{ii} \simeq 0.242$,
  (b) $\bar{S}^z_{ii} \simeq 0.243$, and (c) $\bar{S}^z_{ii} \simeq 0.246$.
  }
  \label{figS3}
\end{figure}

We assume $t_2 /t_1 = 0.9$, where the ground state is the N\'{e}el order.
Figure \ref{figS3} shows the time evolution of the spin correlation function by light irradiation.
When the light has the amplitude $A_0 = 1.0$ and frequency $\omega_{\mathrm{p}} = 30 / t_1$
[Fig. \ref{figS3}(a)], the N\'{e}el order is slightly enhanced while the $120^\circ$ order is suppressed.
When we increase the amplitude to $A_0 = 2.0$ [Fig. \ref{figS3}(b)], the state is switched from
the N\'{e}el order to the $120^\circ$ order after the light irradiation.
When we continue to increase the amplitude to $A_0 = 4.0$ [Fig. \ref{figS3}(c)], the system remains to be the N\'{e}el order, and the corresponding spin correlation is magnified like the result at $A_0 = 1.0$.
These results indicate that the $120^\circ$ order may be induced from the N\'{e}el order by
adjusting the parameter of the external field.

\begin{figure}[tb]
  \centering
  \includegraphics[width=0.8\columnwidth]{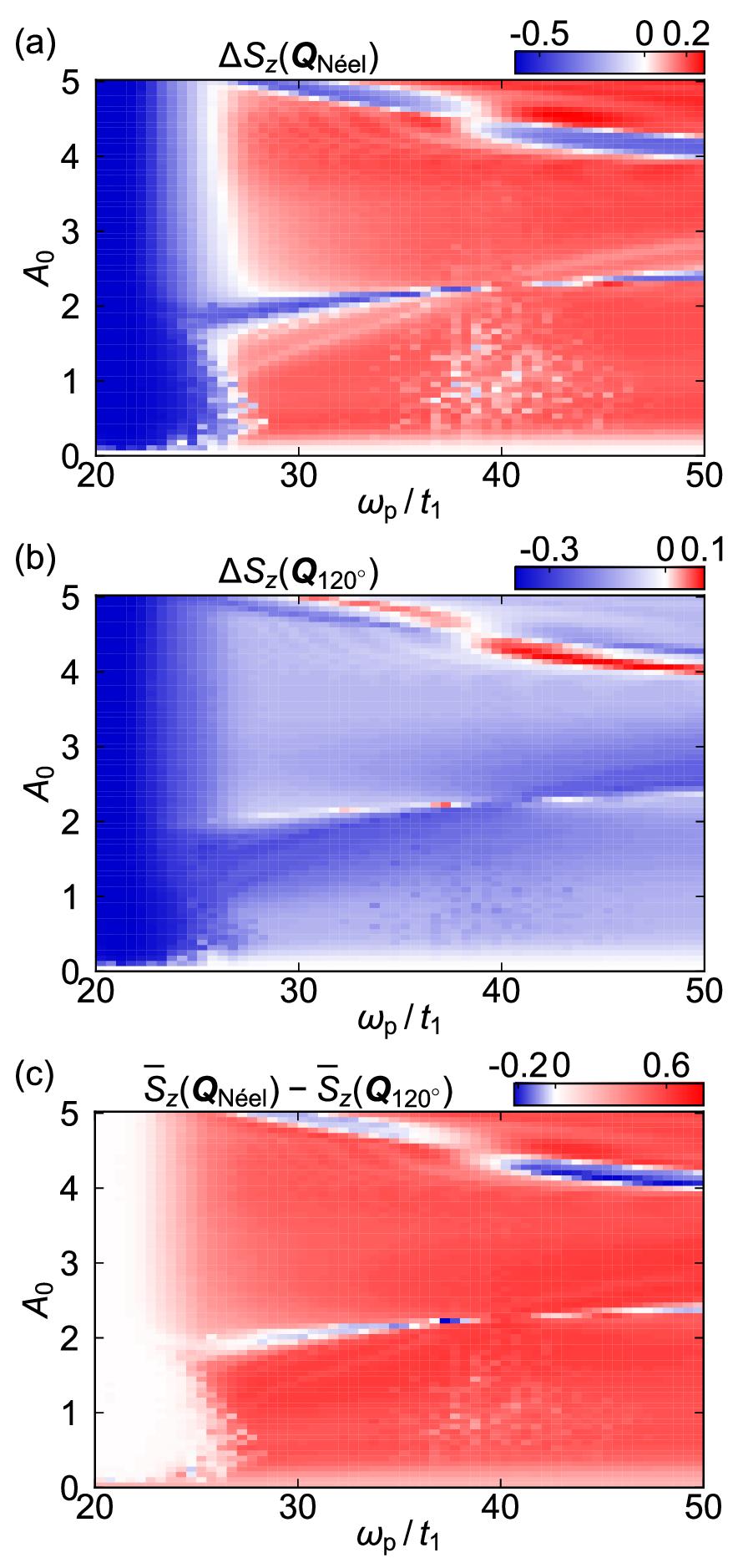}
  \caption{
  Differences in the time-averaged spin correlation function calculated in the parameter space
  $(\omega_{\mathrm{p}}/t_1,A_0)$ (a) $\Delta S_z (\bm{Q}_{\text{N\'{e}el}})$,
  (b) $\Delta S_z (\bm{Q}_{120^{\circ}})$, and
  (c) $\bar{S}_z (\bm{Q}_{\text{N\'{e}el}}) - \bar{S}_z (\bm{Q}_{120^{\circ}})$.
  We assume $t_2/t_1 = 0.9$, i.e., the ground state is the N\'{e}el order.
  }
  \label{figS4}
\end{figure}

As is done in the main text, we also show the differences of the correlation functions
$\Delta S_z (\bm{Q}_{\text{N\'{e}el}})$ and $\Delta S_z (\bm{Q}_{120^\circ})$
in $(\omega_{\mathrm{p}}/t_1,A_0)$ space in Figs.~\ref{figS4}(a) and (b), respectively.
We find that the $120^\circ$ order is enhanced in the region where the N\'{e}el order is suppressed,
and vice versa.
We also show $S_z (\bm{Q}_{\text{N\'{e}el}}) - S_z (\bm{Q}_{120^\circ})$ in Fig.~\ref{figS4}(c).
Thus, the photo-induced magnetic transition from the N\'{e}el order to $120^\circ$ order
can actually occur in the blue region.

\begin{figure}[tb]
  \centering
  \includegraphics[width=0.8\columnwidth]{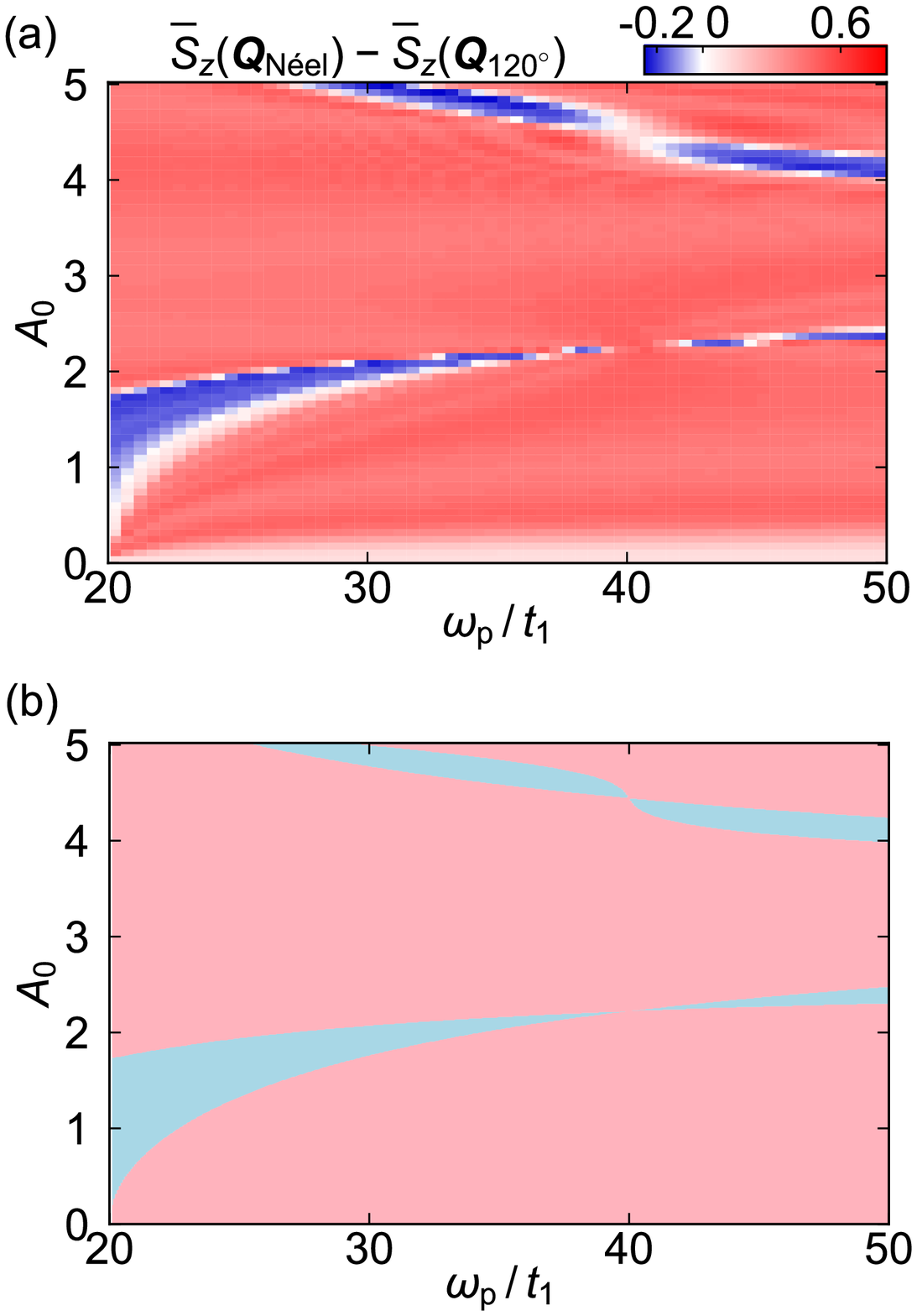}
  \caption{
    (a) Difference $\bar{S}_z (\bm{Q}_{\text{N\'{e}el}}) - \bar{S}_z (\bm{Q}_{120^{\circ}})$
    calculated for the effective Heisenberg model in the parameter space $(\omega_{\mathrm{p}}/t_1,A_0)$.
    (b) Ratio of the Floquet effective exchange interactions, where the region of
    $J_2^{\text{eff}}/J_1^{\text{eff}} < 0.83$
    ($J_2^{\text{eff}}/J_1^{\text{eff}} > 0.83$)
    is indicated by red (blue) color.
  }
  \label{figS5}
\end{figure}

This N\'{e}el-to-$120^\circ$-order phase transition can also be interpreted from the Floquet theory.
The effective Heisenberg-model Hamiltonian derived from the strong-coupling limit is obtained
in the same manner as in the main text.
We again perform the quench dynamics calculation, where the effective exchange interactions
$J_1$ and $J_2$ suddenly change to the Floquet effective exchange interactions
$J^{\text{eff}}_1$ and $J^{\text{eff}}_2$, respectively.
Figure \ref{figS5}(a) shows the difference $S_z (\bm{Q}_{\text{N\'{e}el}}) - S_z (\bm{Q}_{120^\circ})$
after this quench dynamics, the result of which is consistent with the result shown in Fig.~\ref{figS4}(c).
The phase diagram expected from the ratio $J^{\text{eff}}_2 / J^{\text{eff}}_1$ is shown in Fig.~\ref{figS5}(b).
The phase boundary is determined from the line $J^{\text{eff}}_2 / J^{\text{eff}}_1 = 0.83$.
We thus clearly find that the phase diagram obtained by the quench-dynamics calculation
is consistent with the phase diagram determined from the exchange interaction ratio,
implying that the Floquet theory also works in the photo-induced N\'{e}el-to-$120^\circ$-order
phase transition.

\section{Collinear order phase}

\begin{figure}[b]
  \centering
  \includegraphics[width=\columnwidth]{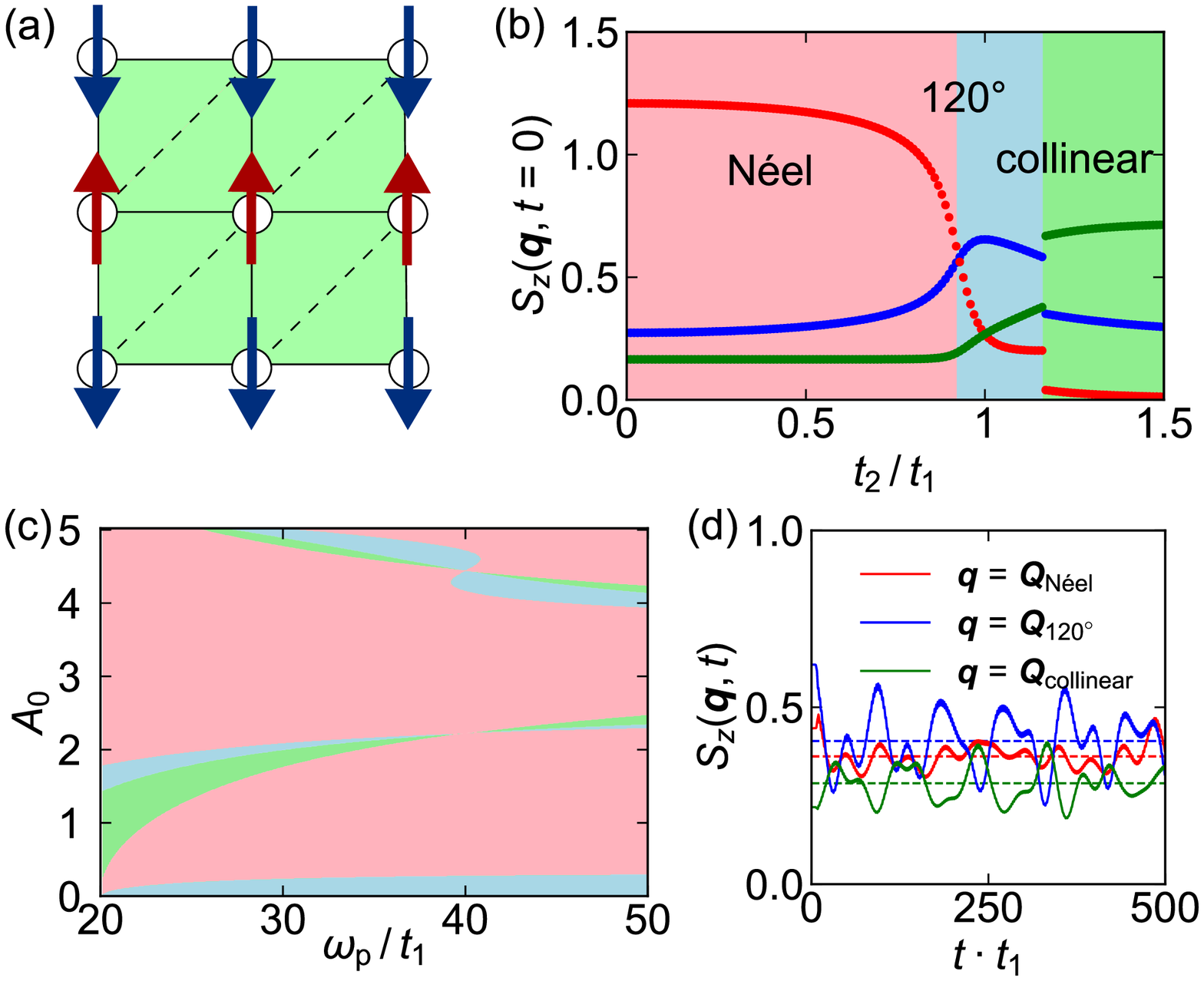}
  \caption{
  (a) Schematic illustration of the collinear order.
  (b) Spin correlation function $S_z (\bm{q}, t = 0)$ calculated for the equilibrium state
  with the N\'{e}el order [$\bm{q} = \bm{Q}_{\text{N\'{e}el}} = (\pi, \pi)$, red line],
  $120^\circ$ order [$\bm{q} = \bm{Q}_{120^\circ} = (2\pi/3, 2\pi/3$), blue line], and
  collinear order [$\bm{q} = \bm{Q}_{\text{collinear}} = (\pi, 0)$, green line], where we assume $U / t_1 = 20$.
  (c) The phase diagram determined from the ratio of the Floquet effective exchange interactions:
  the region of $J_2^{\text{eff}}/J_1^{\text{eff}} < 0.83$ (N\'{e}el state),
  $0.83 < J_2^{\text{eff}}/J_1^{\text{eff}} < 1.53$ ($120^\circ$ state), and
  $J_2^{\text{eff}}/J_1^{\text{eff}} > 1.53$ (collinear state) is indicated by blue, red, and green, respectively.
  The ground state is the $120^\circ$ order ($t_2/t_1 = 0.95$).
  (d) Calculated time evolution of the spin correlation function $S_z(\bm{q},t)$ at
  $\bm{q}=(\pi,\pi)$ (red line), at $\bm{q}=(2\pi/3,2\pi/3)$ (blue line), and at
  $\bm{q}=(\pi, 0)$ (green line), where we assume $A_0 = 1.9$ and $\omega_{\mathrm{p}} = 30$.
  }
  \label{figS6}
\end{figure}

It is known that when the next-nearest-neighbor hoppping integral $t_2$ has a large value,
the magnetic configuration becomes the collinear order illustrated in Fig.~\ref{figS6}(a)
\cite{Mizusaki2006PRB, Yu2010PRB, Misumi2016JPSJ}.
Figure \ref{figS6}(b) shows the spin correlation functions with the wave vectors corresponding
to the N\'{e}el, $120^\circ$, and collinear orders as functions of $t_2 / t_1$ in the ground state.
We find that the system shows the collinear order at $t_2/t_1 > 1.16$, where the correlation
functions have the discontinuity at the transition point.
We believe that this discontinuity is rather artificial because our calculation is restricted
to the small-size cluster, where the finite-size effect is unavoidable.

Figure \ref{figS6}(c) shows the phase diagram including the collinear order in the $(\omega_{\mathrm{p}}/t_1,A_0)$ space.
The phase boundary is determined from the ratio of the effective exchange interactions
obtained from the Floquet theory in the strong coupling limit.
The green region satisfies $J^{\text{eff}}_2 / J^{\text{eff}}_1 > 1.53$, where the collinear order
is expected to emerge.
We calculate the time evolution of the spin correlation functions corresponding to the
N\'{e}el, $120^\circ$, and collinear orders at $t_2 / t_1 = 0.95$ (i.e., the ground state
is $120^\circ$ order) using the parameters corresponding to the green region
($A_0 = 1.9$ and $\omega_{\mathrm{p}} = 30$).
At this point, the ratio of the effective exchange interaction is
$J^{\text{eff}}_2 / J^{\text{eff}}_1 \simeq 3.12$.
However, the spin correlation function of the collinear order never becomes larger than
the other two orders [Fig.~\ref{figS6}(c)], implying that the phase transition from the
$120^\circ$ order to the collinear order by light irradiation does not occur.
We confirm that the quench-dynamics calculation of the Heisenberg model also does
not show such transitions. The realization of this transition remains for a future issue.

\section{Isotropic triangular lattice}

\begin{figure*}[t]
  \centering
  \includegraphics[width=2\columnwidth]{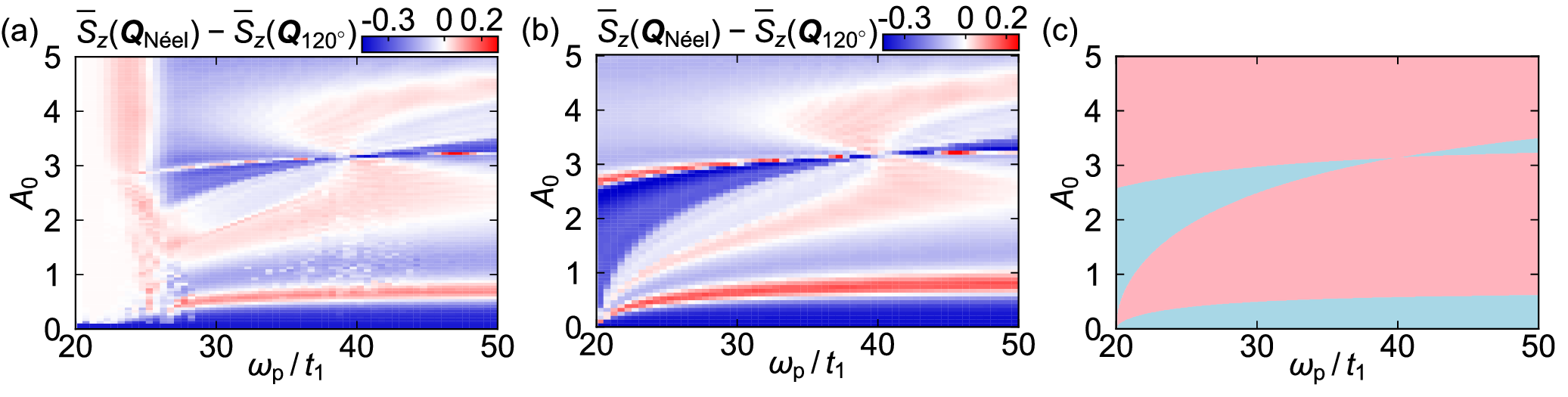}
  \caption{
  (a) Differences in the time-averaged spin correlation function
  $\bar{S}_z (\bm{Q}_{\text{N\'{e}el}}) - \bar{S}_z (\bm{Q}_{120^{\circ}})$
  calculated in the parameter space $(\omega_{\mathrm{p}}/t_1,A_0)$.
  We assume $t_2/t_1 = 0.9$, i.e., the ground state is the N\'{e}el order.
  (b) Difference $\bar{S}_z (\bm{Q}_{\text{N\'{e}el}}) - \bar{S}_z (\bm{Q}_{120^{\circ}})$
  calculated for the effective Heisenberg model.
  (c) Ratio of the Floquet effective exchange interactions: the region of
  $J^{\text{eff}}_2 / J^{\text{eff}}_1 < 0.83$ is indicated by red, while the region of
  $J^{\text{eff}}_2 / J^{\text{eff}}_1 > 0.83$ is indicated by blue.
  }
  \label{figS7}
\end{figure*}

We also consider the most typical case, i.e., an isotropic triangular lattice, where the
next-nearest-neighbor hopping integral satisfies $t_2 = t_1$.
In this case, the ground state is the $120^\circ$ order.

Figure \ref{figS7}(a) illustrates the calculated phase diagram in the $(\omega_{\mathrm{p}}/t_1,A_0)$ space.
The $120^\circ$-to-N\'e{e}l transition occurs in the red region.
This result indicates that the photo-induced magnetic transition can be realized even
if the ground state locates away from the phase boundary.
The phase diagrams obtained from the quench dynamics of the effective-Heisenberg model
in the strong-coupling limit and the effective exchange interaction ratio are shown
in Figs.~\ref{figS7}(b) and \ref{figS7}(c), respectively.
The phase diagrams agree well with each other, indicating that the interpretation by
the strong-coupling limit and Floquet theory also works well in the case of the isotropic
triangular lattice.

\section{Realization in experiments}
We now briefly discuss how to realize the photo-induced magnetic-order switching in experiments, which may be possible because the control of the exchange interactions has actually been succeeded in iron oxides \cite{Mikhaylovskiy2015NC}.

One of the candidate materials may be $\kappa$-type organic salts, which are well described by the anisotropic triangular-lattice Hubbard model \cite{Hotta2012Crystals}. $\kappa$-(BEDT-TTF)$_2$-Cu[N(CN)$_2$]Cl, in particular, has the N\'{e}el order at low temperatures \cite{Lefebvre2000PRL}, implying that the N\'{e}el-to-120$^\circ$-order transition can occur by photo-irradiation. However, because this transition can be induced only in a somewhat restricted parameter region, as discussed in Sec. \ref{Sec:Neel-to-120-order}, the realization of the photo-induced phase transition in this material seems not to be easy. Moreover, the fact that the organic salts with the $120^\circ$ order have not been discovered so far seems to suggest an additional difficulty in its realization.

Another candidate material includes inorganic compounds such as Cs$_2$CuCl$_4$ \cite{Coldea2002PRL} and Cs$_2$CuBr$_4$ \cite{Ono2003PRB}, which exhibit the $120^{\circ}$ magnetic order behavior at low temperatures. Here, to avoid a heating effect by the strong persistent periodic field, we may instead observe the transient dynamics by a pump-probe measurement. The time-dependent resonant inelastic x-ray scattering spectroscopy \cite{Dean2016NM} may also be used to investigate the transient spin structure.
Compounds with the isotropic triangular lattice, such as
CsCuCl$_3$ \cite{Adachi1980JPSJ},
Ba$_3$CoSb$_2$O$_9$ \cite{Shirata2012PRL, Zhou2012PRL, Ma2016PRL, Ito2017NC}, and Li$_2$InMo$_3$O$_8$ \cite{Haraguchi2015PRB, Iida2019SR},
which host the $120^\circ$ order,
may also be candidates for the photo-induced magnetic phase transition.
However, these inorganic compounds have relatively small $T_{\mathrm{N}}$,
implying that their $120^\circ$ order may be fragile against thermalization by excited electrons.

A different route to realizing the photo-induced phase transition may be to use the ultracold atomic gases
in optical lattices \cite{Eckardt2017RMP}. The anisotropic triangular lattice with ultracold bosons
has already been realized experimentally \cite{Eckardt2010EPL, Struck2011Science},
so that the expected long-range orders may be observed by varying the nearest- and
next-nearest-neighbor coupling elements independently.
The realization of ultracold fermions in the anisotropic triangular lattice may also be feasible.
Here, a proper driving force \cite {Messer2018PRL, Sandholzer2019PRL} corresponding to the
external periodic field discussed in the present paper can be introduced, which may possibly lead
to the magnetic phase transitions of the system.
